# MVP2P: Layer-Dependency-Aware Live MVC Video Streaming over Peer-to-Peer Networks

Zhao Liu[1], Niall Murray[2], Brian Lee[1], Enda Fallon[2] and Yuansong Qiao[1*]

zliu@research.ait.ie, nmurray@research.ait.ie, blee@ait.ie, efallon@ait.ie, ysqiao@research.ait.ie

[1]Software Research Institute, Athlone Institute of Technology, Athlone, Co. Westmeath, Ireland

[2]Department of Electronics & Informatics, School of Engineering, Athlone Institute of Technology, Athlone, Co. Westmeath, Ireland

*Abstract*—**Multiview video supports observing a scene from different viewpoints. The Joint Video Team (JVT) developed H.264/MVC to enhance the compression efficiency for multiview video, however, MVC encoded multiview video (MVC video) still requires high bitrates for transmission. This paper investigates live MVC video streaming over Peer-to-Peer (P2P) networks. The goal is to minimize the server bandwidth costs whist ensuring high streaming quality to peers. MVC employs intra-view and inter-view prediction structures, which leads to a complicated layer dependency relationship. As the peers' outbound bandwidth is shared while supplying all the MVC video layers, the bandwidth allocation to one MVC layer affects the available outbound bandwidth of the other layers. To optimise the utilisation of the peers' outbound bandwidth for providing video layers, a maximum flow based model is proposed which considers the MVC video layer dependency and the layer supplying relationship between peers. Based on the model, a layer dependency aware live MVC video streaming method over a BitTorrent-like P2P network is proposed, named MVP2P. The key components of MVP2P include a chunk scheduling strategy and a peer selection strategy for receiving peers, and a bandwidth scheduling algorithm for supplying peers. To evaluate the efficiency of the proposed solution, MVP2P is compared with existing methods considering the constraints of peer bandwidth, peer numbers, view switching rates, and peer churns. The test results show that MVP2P significantly outperforms the existing methods.**

## 1. INTRODUCTION

MULTIVIEW Video Coding (MVC) has been widely recognized as one of the key technologies to serve multiview video applications, e.g. 3D televisions and free viewpoint televisions [1] [2]. Multiview video consists of a number of views captured from multiple cameras (normally placed very close to one another) to cover different portions of a scene. The Joint Video Team (JVT) developed H.264/MVC [2] [3] (in short: MVC) based on the widely deployed H.264/Advanced Video Coding (AVC) [4]. In addition to using the temporal intra-view predictions as defined in H.264/AVC, the spatial redundancies between views are reduced by using a new inter-view prediction method [3]. This enhances MVC compression efficiency but introduces a complex layer dependency relationship.

Peer-to-Peer (P2P) streaming technologies have been widely adopted to deliver large volumes of multimedia contents to a large-scale of users at low infrastructural costs [5] [6]. When streaming MVC video in P2P networks, a fundamental problem is in allocating the outbound bandwidth of peers to supply different layers. MVC video has a mesh type layer dependency structure. When a peer selects a view to watch, it may need to download only a part of the video layers to decode the view. Therefore, there may be many copies for some layers and very few copies for some other layers in the network. When a peer supplies layers to other peers, all the layers share the same set of outbound network links. Consequently, if some layers occupy a large portion of the peer's outbound bandwidth, the supplying speeds for the other layers have to be reduced. When a peer plans to download a certain layer (or chunks), if it cannot find any peers that can supply the content, it has to resort to the server. The challenge therefore is how to globally coordinate the outbound bandwidth allocation amongst all the peers so that the peers' outbound bandwidth can be maximally utilized.

Layered video streaming over P2P networks has been extensively studied, e.g. for AVC-type video streaming [7] [8] or for SVC (Scalable Video Coding) video streaming [9] [10]. The research focus has been mainly in scalable (or adaptive) streaming, i.e. adjusting the streamed video layers according to the bandwidth conditions. Exploring the video layer dependency to optimise the global streaming performance has not been considered in the existing works. The essential difference between AVC/SVC video streaming and MVC video streaming is in their application scenarios, i.e. AVC/SVC is for one view whereas MVC is for multiple views. Therefore the design emphasis is different even though the video encoding structures are similar (especially SVC and MVC). In AVC/SVC streaming, the lower layers in a video stream are deemed to be more important than the higher layers to ensure successful decoding (or smooth playbacks). In MVC video streaming, when a view is selected by the user, all the layers that are required to decode the view have to be downloaded.

* Corresponding author

Current works on streaming MVC video over P2P networks are very limited. The existing solutions generally employ the traditional P2P video streaming approaches, e.g. [11] [12] [13]. In [11] [12], each view is streamed over an independent overlay network. The peers observing different views cannot share all their common video layers. In [13], each peer requests chunks in a timely manner by employing a sliding window scheduling mechanism. The chunks are randomly pulled from any neighbouring peers. The above approaches lack awareness of the heterogeneous layer availabilities amongst peers, and fail to optimally utilize peers' outbound bandwidth. Section 3.2 will provide a detailed analysis on the problems of traditional methods.

This paper proposes MVP2P, a live MVC video streaming system over P2P networks. The main application scenario considered is free viewpoint televisions. In this scenario, peers choose a view to watch and all the required layers for decoding the chosen view must be obtained within a strict time constraint. The goal of the design is to guarantee that all the peers obtain their required layers whilst simultaneously minimizing the server bandwidth cost. The following factors are considered in the design: (1) System scalability (e.g. server bandwidth costs); (2) User's QoE (Quality of Experience), in particular video playback smoothness, start-up delays and view switching delays; (3) The mesh-shaped MVC layer dependency structure, which complicates the design; (4) MVC's application scenarios. Unlike AVC or SVC video streaming, layer switching in MVC is caused by both viewpoint switching and quality adaptation. Furthermore, quality adaptation in an MVC video streaming is constrained by the peers' chosen views. (5) Network dynamics. In addition to peer churns, view switching happens frequently in free viewpoint television scenarios.

The contribution of this paper is summarized as follows.

(1) A maximum flow based model is proposed to capture the MVC layer dependency and the layer supplying relationship amongst peers. The model provides a guideline for allocating peers' outbound bandwidth to supply different layers. It can maximize the utilisation of peers' outbound bandwidth and consequently minimize the server bandwidth costs.

(2) Based on the model, a layer dependency aware MVC video streaming method over a BitTorrent-like P2P network is proposed, i.e. MVP2P. The above maximum flow based model is applied to both the chunk requesting and supplying processes in peers. For the requesting process, a new chunk scheduling strategy and a new peer selection strategy are proposed to ensure that the chunk requesting process complies with the maximum flow based model. For the supplying process, a new outbound bandwidth scheduling algorithm is proposed to reduce the chunk dissemination delays. The evaluation results show that the streaming performance of MVP2P is very close to the theoretical optimal values calculated from the maximum flow based model and it is significantly better than the existing solutions. To the best of our knowledge, this is the first piece of work on optimizing MVC video streaming over P2P networks by considering the MVC layer dependency relationships.

The rest of this paper is organized as follows. Related work is summarized in Section 2. In Section 3, the problems in streaming MVC video are discussed and the high level design concept is presented. The detailed MVP2P design is described in Section 4. Section 5 evaluates the performance. Finally, we conclude the paper with Section 6.

## 2. RELATED WORK

The related work is classified as non-scalable video streaming, scalable video streaming and multiview video streaming.

**Non-scalable video streaming:** In [17], Pai et al., propose a random chunk scheduling algorithm. Chunks are randomly pulled from supplying peers, which may waste peer bandwidth to some extent [6]. In [18], Peterson et al., use a coordinator to efficiently direct the peer bandwidth allocation amongst different randomly-organized swarms. In [14] [16], the tit-for-tat strategy (peer selection strategy) is applied to incentivize peers contributing bandwidth to the network. For the chunk scheduling strategy, the rarest-first strategy is used to enhance the chunk availabilities in the network and the sequential chunk scheduling strategy is used to smooth the video playback. In [15], peer selection strategies for video-on-demand streaming are studied. In [19], several random chunk scheduling strategies are compared. Various other P2P streaming technologies investigate the heterogeneous peer properties and the underlying network conditions, e.g., locality together with efficient link bandwidth utilization (P4P [20] and ALTO protocol [21]) and load balancing [22] [23]. In non-scalable video streaming, each peer retrieves the same video stream, whereas in MVC video streaming, different peers may require and supply different video layers. This paper focuses on the impact of the heterogeneous layer availabilities on the performance of MVC video streaming, and proposes an effective approach to utilize the peers' outbound bandwidth.

**Scalable video streaming:** In scalable video streaming over P2P networks, peers can adjust received video quality according to current network conditions. In [10], Qiao et al., propose a scalable video streaming system over P2P networks (SVDN). In [24], Goktug Gurler et al. propose an adaptive sliding window mechanism to maximize the chunks for improving a peer's received video quality. There are many other research works [7] [8] [9] investigating on maximizing the received video qualities for the global peers. In [7], Cui et al. propose a cost-effective greedy algorithm for AVC-type video streaming. The dependency relationship between layers is assumed to be linear. The algorithm recursively utilizes the bandwidth of peers with the smallest number of layers. In [8], the chunk scheduling problem is regarded as the knapsack problem. Several chunk scheduling strategies (e.g. greedy approximation algorithm and dynamic programming) for layered video streaming over P2P networks are evaluated. In [25], Shen et al., investigate the importance of the chunks for improving video quality. The chunks are assigned with different weights and scheduled in the order of the weights. In [9], Mokhtarian et al., propose to use peer outbound bandwidth to provide the lower layers first in SVC video streaming, and then they use approximation algorithms for allocating servers' bandwidth resources to peers in order to achieve higher SVC video streaming qualities.

All the above approaches are designed for the video quality adaptation applications where all the video layers are for one viewpoint. The peers can reduce the received video quality when the bandwidth is insufficient or vice versa. MVC has a similar layer dependency structure to SVC. However, the application scenario is different. In MVC video streaming, different video layers may belong to different viewpoints. Once a viewpoint is chosen by the user, the video layers for decoding the viewpoint have to be streamed to the user. Directly applying the above approaches to MVC streaming may violate users' preferences (e.g. unable to decode the user selected view) or cause a suboptimal streaming performance because the video layer sharing amongst peers are not coordinated globally.

**Multiview video streaming:** There are two main methods to encode multiview video [26]: (1) dependent encoding using MVC or multiview coding extension of H.265/High Efficiency Video Coding (HEVC) [27]; (2) simulcast encoding in which each view is independently encoded using AVC, SVC or HEVC [28].

In traditional client/server based architectures, various MVC video streaming approaches have been proposed, e.g. [29] [30] [31]. In [29] [30], Liu et al investigate the effects of network conditions on the streamed MVC video qualities and propose a NALU interleaving scheme for real-time MVC video streaming. In [31], Terence et al propose a scalable Free-Viewpoint TV broadcasting architecture over Long-Term Evolution cellular networks, which focuses on the transmission format to reduce the bandwidth consumption.

For MVC video streaming over P2P networks, the existing works mainly focus on overlay network construction and adaptive streaming. In [12], Kurutepe, et al. propose a multi-tree P2P overlay network to stream an MVC video. Each view is streamed over an independent P2P multicast tree. Each peer is constrained to join an independent overlay network for a view and contribute uplink bandwidth to that network. The peers from different overlay networks cannot share their common layers. In [11], Chen et al. propose a two-layered overlay network. The users observing the same view are organized into an overlay network called intra-view overlay. Users of different intra-view overlays are organized together to construct a global overlay network called cross-view overlay. The peers observing different views only share the base layer and do not share the other common layers. In [13], adaptive MVC video streaming over BitTorrent-like P2P networks has been studied. Each peer maintains a sliding window for managing its chunks of each view. The peer can dynamically discard views as per their network conditions. The missing chunks in the sliding window are requested in a timely manner. The chunks are randomly picked from any available supplying peers. In this paper, we focus on the chunk scheduling strategy and peer selection strategy by considering MVC layer dependency.

There are many existing works that investigate on simulcast multiview video streaming over P2P network. In [32], Ding et al., propose a clustering-based overlay network which enables a peer to select neighbouring peers with overlapped interests. They propose chunk scheduling algorithms to allow peers to select chunks for maximizing their 3D video qualities in dynamic networks. In [2] [26] [33] [34], the adaptive multiview video streaming over BitTorrent-like P2P networks is investigated. They consider the scenario of 3D televisions. Similar to scalable video streaming, lower layers are assigned higher important levels. In [35], Ren et al. investigate the problem of sharing the peers' reference views for the virtual view synthesis in live free viewpoint video streaming scenarios. They propose an algorithm to help peers choose their reference views and neighbouring peers based on minimizing the total streaming costs (e.g. network reconfiguration cost). In [36], Yuan et al. assume that each peer has two reference views. Each peer simultaneously joins in two overlay networks of its reference views so that the peers can efficiently share their reference views. In above streaming approaches, multiview videos are not encoded using MVC. In this paper, we investigate the effects of MVC layer dependency on MVC video streaming performance.

## 3. PROBLEM DESCRIPTION

### 3.1 *MVC Encoding Scheme and Layer Dependency Relationship*

In MVC, three types of views are defined based on the use of inter-view predictions, as illustrated in Fig. 1. In Fig. 1, *S0, S1, ...* and *T0, T1, ...* represent parallel views and temporal positions of pictures respectively.

(1) **Base view.** It is coded independently using only intra-view prediction, e.g. View *S0* in Fig. 1.

(2) **Predicted View.** It is encoded based on a previous reference view, e.g. View *S2* and *S4* in Fig. 1. However, only the picture at the boundary of a Group of Picture (GOP) is predicted from the preceding reference views.

(3) **Bi-predicted View.** It is predicted based on both the previous view and the next view. All pictures are coded using inter-view prediction and intra-view prediction, e.g. View *S1* and *S3* in Fig. 1.

All the views are jointly coded to form a single MVC bit-stream and organized into network friendly Network Abstraction Layer Units (NALUs). NALU is specified to encapsulate the coded picture information and provides header information, e.g. view identifier (VID) and temporal layer identifier (TID), for media transmission or storage. The NALUs with the same VID and TID are called a layer. A layer represents a set of NALUs in the hierarchical prediction structure [1].

For a specific MVC prediction structure, there is a corresponding layer dependency relationship. Fig. 2 shows the MVC layers dependency relationship with respect to Fig. 1. A layer is identified by a VID and a TID, e.g., layer $L_{1,2}$ denotes the layer with VID equal to 1 and TID equal to 2. The maximum TID is 3 because the GOP size is 8 in Fig. 1, which is calculated based on the formula $log_2 N$ ($N$ is the GOP size) [2]. As we can see, except layer $L_{0,0}$ (the base layer which is decoded independently), all the others layers are encoded based on their reference layers. In this paper, we introduce new notations – **higher layer** and **lower layer**. If layer $B$ is encoded based on layer $A$, we say that layer $B$ is a higher layer of layer $A$ and conversely layer $A$ is a lower layer of layer $B$. MVC layer dependency has a mesh-like structure. For example, $L_{0,0}$, $L_{2,0}$ and $L_{1,0}$ are in a linear relationship whereas $L_{0,1}$, $L_{2,1}$ and $L_{1,1}$ form a tree. $L_{1,0}$ is encoded based on $L_{0,0}$ and $L_{2,0}$ for sure. However, as $L_{2,0}$ also depends on $L_{0,0}$, it is not necessary to connect $L_{0,0}$, and $L_{1,0}$ in Fig. 2.

The NALUs are arranged in Decoding Order Number (DON) within an MVC bit-stream [2]. Accordingly, we specify a new notation – **layer DON** – to describe the relative locations between the layers in an MVC bit-stream. The layers with lower TIDs are arranged earlier. For the layers with the same TID, they are arranged based on their VIDs. The decoding order for VIDs in Fig. 1 is shown as follows: $S_0$, $S_2$, $S_1$, $S_4$ and $S_3$.

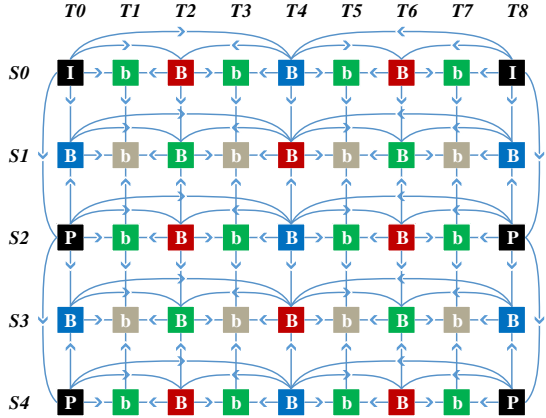

Fig. 1 Typical MVC prediction structure [3]

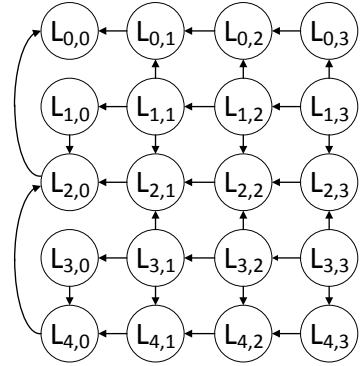

Fig. 2 A typical MVC layer dependency relationship

### 3.2 *Problem Statement*

Let us study the problem of MVC video streaming with a simple example. Fig. 3 demonstrates the problems of applying a traditional greedy algorithm for streaming layered video [7] in MVC video streaming scenarios. In the greedy algorithm, a peer always prefers to request chunks from the available peers with the smallest number of layers. This algorithm is based on the assumption that peers with higher layers can supply lower layers. Therefore, utilizing the outbound bandwidth of peers with lower layers can save the bandwidth of peers with higher layers, and consequently increasing the opportunities of supplying higher layers using peers bandwidth instead of the servers'. This assumption is true for video codecs with a linear dependency relationship, e.g. AVC, but it does not fit for MVC codecs because of its mesh-like layer dependency structure (i.e. decoding a layer does not always require the layers with smaller DONs, e.g. Fig. 2).

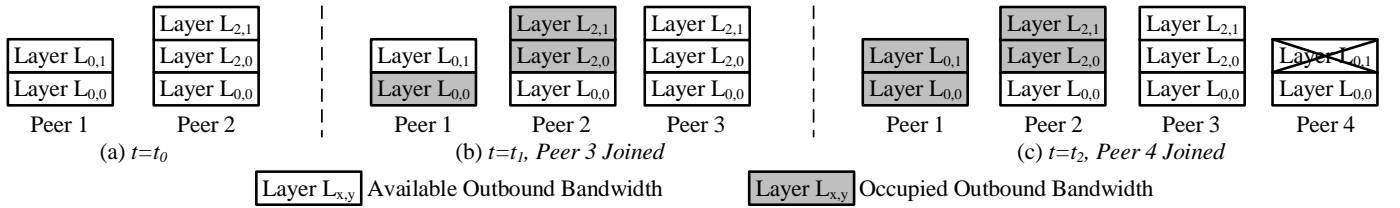

Fig. 3 An example of the greedy bandwidth scheduling algorithm

In this example, we have following assumptions. The MVC video has a layer dependency as shown in Fig. 2. The bitrate of each layer is the same. For each peer, the inbound bandwidth is just sufficient to maintain continuous downloading of all its required layers, and the inbound bandwidth and outbound bandwidth are equal. Each peer observes a view at a certain temporal layer (in short: **observing layer**). Each peer only needs to download the required layers for decoding its observing layer.

At the starting point (Fig. 3a), two peers exist in the network: Peer 1 watches on layer $L_{0,1}$ and Peer 2 watches on layer $L_{2,1}$. According to Fig. 2, layer $L_{0,0}$ is a lower layer of layer $L_{0,1}$. Layer $L_{0,0}$ and $L_{2,0}$ are lower layers of layer $L_{2,1}$. Peer 1 has just enough outbound bandwidth to provide layer $L_{0,0}$ and $L_{0,1}$. Peer 2 has just enough outbound bandwidth to provide layer $L_{0,0}$, $L_{2,0}$ and $L_{2,1}$.

In Fig. 3b, a new peer Peer 3 joins the network and watches on layer $L_{2,1}$. According to the greedy algorithm, Peer 1 is selected to contribute layer $L_{0,0}$ to Peer 3 because Peer 1 has less number of layers than the Peer 2. Afterwards Peer 2 is selected to contribute layer $L_{2,0}$ and layer $L_{2,1}$ to Peer 3.

In Fig. 3c, another new peer Peer 4 joins the network and watches on layer $L_{0,1}$. As Peer 1 has the smallest number of layers than others, Peer 1 is selected as the first supplying peer for Peer 4. As Peer 1 is providing layer $L_{0,0}$ for Peer 3, the outbound bandwidth of Peer 1 is only enough to provide layer $L_{0,0}$ for the Peer 4. Peer 4 cannot obtain Layer $L_{0,1}$ from other peers since only Peer 1 contains layer $L_{0,1}$. In order to watch on layer $L_{0,1}$, Peer 4 must obtain layer $L_{0,1}$ from the servers, which will increase server bandwidth costs.

According to the MVC layer dependency structure, MVC layers are unequally important from the decoding perspective, e.g. the base layer $L_{0,0}$ is required for decoding all the layers whereas layer $L_{0,1}$ is required for decoding some layers in Fig. 2 and Fig. 3. This leads to unequal distribution of layers over the network, e.g. layer $L_{0,0}$ exists in all peers whereas layer $L_{0,1}$ only exists in Peer 1 in Fig. 3. The above example has already shown that the streaming method (e.g. the greedy algorithm) directly affects the server performances and the streamed video qualities to end users. This paper investigates the potential of maximally utilizing the peers' outbound bandwidth to supply layers for global peers so that the servers' bandwidth can be minimized.

### 3.3  Problem Modelling and Formulation

The optimization objective of MVP2P is to minimize server bandwidth costs and also improve streamed qualities for end users (e.g. reducing chunk downloading delays to ensure a smooth playback). The two optimization objectives sometimes contradict each other. For example, to minimize the server bandwidth costs, the peers' outbound bandwidth should be maximally utilized. Hence, video chunks should be always relayed from one peer to another. This produces chunk dissemination delays among peers and may reduce the streamed qualities to peers (e.g. playback stall). Conversely, to ensure the peers' smooth playbacks, chunks must be delivered to all peers before the decoding deadlines. The chunk dissemination delays that exceed the deadline is not allowed. The server bandwidth may be required for reducing the chunk dissemination delays. In this paper, the main optimization objective is set to minimize the server bandwidth costs and the chunk dissemination delays are chosen as constraints.

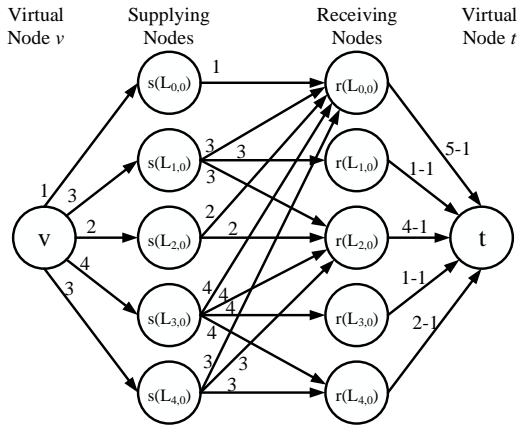

Fig. 4 The maximum flow model for peer based MVC video streaming

A maximum flow model (Fig. 4) is constructed to depict the MVC layer dependency relationships and the layer supplying relationship between peers. The model will be used as a guideline for peers to request or supply chunks. The following presumptions are required as the basis of the model.

(1) Each peer only watches one view at each moment. This is applicable to the scenario of the first phase of free viewpoint televisions [37]. Future free viewpoint televisions can support multiple views [27]. The proposed maximum flow model can be easily extended to support this scenario, which will be discussed in the end of this section.

(2) Each peer chooses a view (and the respective temporal layer) either based on user's preference or inbound bandwidth limitations. The decision process of choosing the view and dynamic video quality adaptation (i.e. switching between temporal layers) is outside of the scope of the paper. We assume that, once each peer has chosen a view, the peers' inbound bandwidth is sufficient to download to required video layers.

The flow graph is shown in Fig. 4 (based on the layer structure in Fig. 2). For simplicity, the graph only shows the lowest layer of each view in Fig. 2. There are three types of nodes in the graph:

(1) Supplying node $s(L_{x,y})$, which represents the peer set watching on (or intending to watch on) a specific layer, e.g. $s(L_{0,0})$ represents all the peers watching on view 0 with temporal layer ID equal to 0 (Layer $L_{0,0}$).

(2) Receiving node $r(L_{i,j})$, which represents the peer set requesting a specific layer, e.g., $r(L_{0,0})$ donates all the peers requesting Layer $L_{0,0}$. An edge is directed from a supplying node to a receiving node if a supplying node can provide the layer of a receiving node based on the MVC layer dependency structure (e.g. Fig. 2). For example, an edge is directed from $s(L_{1,0})$ to $r(L_{0,0})$ because the peers watching on layer $L_{1,0}$ can potentially provide layer $L_{0,0}$. The edge capacity represents the outbound bandwidth of a supplying node, i.e. the total peers' outbound bandwidth of the peer set.

(3) Virtual node. The virtual node "$v$" is the source node in the graph. It is a virtual supplier in the network. It directs an edge to each supplying node. The edge capacity $c(v,\ s(L_{x,y}))$ represents the total outbound bandwidth of the supplying node $s(L_{x,y})$. The virtual node "$t$" is the sink node in the graph. It is a virtual demander. An edge is directed from each receiving node to it. The edge capacity $c(r(L_{i,j}),t)$ represents the total required bandwidth for a specific layer $r(L_{i,j})$ in the network (after reducing one times of the layer bit rate). Let $R(L_{i,j})$ denote the bitrate of layer $L_{i,j}$ and suppose $n$ peers require layer $L_{i,j}$ in the network. $c(r(L_{i,j}),t)=(n-1)\times R(L_{i,j})$ as the servers need to provide at least 1 copy of each layer to the peers.

For a particular layer, it can be cooperatively provided by multiple supplying peers. The goal is to maximize the uploading bit rates of the peers for providing layers and consequently minimize the servers' bandwidth costs. Let us use $\{s(L_{0,0}),....,S(L_{u,v})\}$ to denote the supplying nodes and use $\{r(L_{0,0}),...,r(L_{u,v})\}$ to denote the receiving nodes. Let $u(s(L_{x,y}),\ r(L_{i,j}))$ denote the flow value of the flow going through the supplying node $s(L_{x,y})$ and the receiving node $r(L_{i,j})$. The optimization goal can be formalized as:

$$\text{maximize} \sum_{r(L_{i,j})=r(L_{0,0})}^{r(L_{u,v})} \sum_{s(L_{x,y})=s(L_{0,0})}^{s(L_{u,v})} u(s(L_{x,y}),r(L_{i,j})) \tag{1}$$

We state the problem as routing as much flow as possible from source $v$ to sink $t$. If we consider that the minimum delivery unit is a chunk (or a block in some implementations), i.e. the chunks cannot be divided during transmission and forwarding, this problem is essentially a single source unsplittable flow problem which is NP-complete [7]. If we consider the flows are splittable, this is known as the maximum flow problem [38]. This paper focuses on splitting the outbound bandwidth of the peers while supplying the chunks of different layers, and chooses the maximum flow model as an approximation to NP-complete problem.

After calculation of the maximum flow model, $u(v,\ s(L_{x,y}))$ will be the total bandwidth that peer set $s(L_{x,y})$ will contribute for supplying layers, $u(s(L_{x,y}),\ r(L_{i,j}))$ will be the outbound bandwidth that peer set $s(L_{x,y})$ needs to contribute to the provision of layer $L_{i,j}$, $u(r(L_{i,j}),\ t)$ will be the total bandwidth that all the peers can contribute to layer $L_{i,j}$, and the server bandwidth consumption for layer $L_{i,j}$ (denoted by $Q_s(L_{i,j})$) will be

$$Q_s(L_{i,j}) = c(r(L_{i,j}),t) - u(r(L_{i,j}),t) + R(L_{i,j}) = n \times R(L_{i,j}) - u(r(L_{i,j}),t)$$

$$subject\quad to\quad c(r(L_{i,j}),t) \geq 0 \tag{2}$$

In the case that each peer may observe on multiple views, the minor modification is that we can organize the peers which observe the same layers as a supplying node in Fig. 4. This scenario will not be considered in the paper.

### 3.4 Chunk Dissemination amongst Peers

The above maximum flow model provides a solution that specifies the outbound bandwidth contribution of each peer set for supplying each MVC video layer and the correspondent minimum server bandwidth consumptions. However, the model does not specify how the chunks are actually disseminated amongst peers, and therefore we need to construct a chunk dissemination scheme that complies with the model. In MVP2P, the basic concept is to divide the outbound bandwidth of a peer based on the maximum flow model while supplying the MVC video layers and then utilizing the single layer based P2P streaming technologies to disseminate the chunks of each layer. The chunk dissemination process follows two key principles, i.e. increasing the chunk diversities amongst peers so that peers can share the chunks efficiently, and reducing the chunk dissemination delays (from the creation time of the chunk to the time that all the peers that demand the chunk have received it).

In [46], the authors provide a theoretical analysis to P2P streaming of single layered video and proof that it is feasible to fully utilize peers' outbound bandwidth. The basic idea is to divide a video stream into separate sub-streams and then the servers transmit different sub-streams to different peers. Afterwards, the peers exchange these sub-streams by utilizing their outbound bandwidth. As a video stream cannot be arbitrarily divided (the minimum transmission unit is a chunk), this paper employs this concept and uses heuristic methods to assign sub-streams to peers, e.g. the rarest-first method. Detailed discussions will be presented in Section 4.

There are many streaming approaches (e.g. tree based streaming) for chunk dissemination amongst peers. Snow-ball streaming [45] exhibits an excellent delay performance. It sorts the peers based on the outbound bandwidth in descending order. The server pushes a new chunk to the peer with the largest bandwidth. Afterwards, in each round, the peers that have the chunk will push the chunk to other peers with the largest bandwidth first. This requires a strict chunk dissemination order amongst peers, which is against the characteristics of a BitTorrent-like streaming system where the peers are randomly organized and each peer autonomously decides which chunks should be requested from other peers.

In this paper, we employ a pull based Snow-ball approach, i.e. the concept of the parallel Snow-ball approach is adopted but its strict chunk dissemination order among peers is not followed. In MVP2P, the peers autonomously decide what chunks to be requested from a supplying peer based on the maximum flow model, and the peers may request many chunks at the same time. Unlike the original parallel Snow-ball approach, each supplying peer is allowed to provide different chunks at the same time. The key concept for requesting and supplying chunks are given below. The detailed design will be discussed in Section 4.

As each peer needs to request chunks and can supply chunks simultaneously, it belongs to one supplying node and a number of receiving nodes at the same time in the maximum flow model. While requesting chunks, a peer needs to consider its duty in supplying layers while acting as a supplying node. While supplying chunks, a peer needs to consider its contribution to each specific layer, as well as the duty of the requesting peers when they serve as supplying nodes.

According to the maximum flow model, for each receiving node $r(L_{i,j})$, we can obtain the outbound bandwidth that each supplying node $s(L_{x,y})$ allocates for providing layer $L_{i,j}$. Consequently, for each layer $L_{i,j}$, we can calculate the proportion $\rho_{s(L_{x,y})}$ of the outbound bandwidth that each supplying node $s(L_{x,y})$ contributes to the layer. When a peer requests chunks, it preferentially requests the layers that it needs to supply more to the network. For a specific layer $L_{i,j}$, the peer chooses the supplying peers based on the probabilities derived through the above proportion ($\rho_{s(L_{x,y})}$), i.e. the requests are more likely sent to the supplying nodes with higher contributions to the specific layer. When a peer acts as a supplying node, it divides its outbound bandwidth for each layer based on the above proportion ($\rho_{s(L_{x,y})}$). It preferentially delivers the chunks to the receiving peers with higher contributions to this specific layer. This enables that the peers' outbound bandwidth utilization complies with the maximum flow model in general (as shown in the test section). More details of chunk dissemination among peers are presented in the next section.

## 4. MVP2P DESIGN

### 4.1 *Design Overview*

In a BitTorrent-like P2P streaming system, the video content is encapsulated into chunks for transmission. When a peer acts as a receiving peer, the chunk scheduling strategy decides the downloading orders of the chunks and the peer selection strategy decides the supplying peers for downloading each chunk. When a peer acts as a supplying peer, the supplying peer bandwidth-scheduling algorithm decides how to allocate the outbound bandwidth to send chunks to the receiving peers. They all have impact on the peers' outbound bandwidth utilization. MVP2P proposes novel algorithms for the above process based on the maximum flow model introduced in Section 3. The design principle is to minimize the server bandwidth consumption while guaranteeing the live streaming quality of each peer such as high video qualities, smooth video playbacks, short start-up delays, and short view switching delays.

#### 4.1.1 *Basic Concept in MVP2P Design*

The MVP2P method uses the maximum flow model to calculate the peers' outbound bandwidth allocation for providing each layer. Thereafter, the result of the maximum flow model forms a guideline for both the receiving peers and the supplying peers. For the receiving peers, a new chunk scheduling strategy is proposed to improve chunk availabilities in the network, i.e. increase the chunk diversities in the supplying nodes in the maximum flow model so that the outbound bandwidth of peers can be utilized efficiently. A new peer selection strategy is proposed which enables the peers' outbound bandwidth utilization to comply with the maximum flow model. For supplying peers, a supplying peer bandwidth-scheduling algorithm is proposed to reduce the chunk dissemination delays. Referring to the parallel Snow-ball approach, the above algorithm preferentially delivers each chunk to the peers that allocate more outbound bandwidth for providing the chunk according to the result of the maximum flow model. To ensure the streaming qualities of peers, the chunks are assigned with different importance levels. As the chunks that the peers require for maintaining smooth video playback are more important, the receiving peers preferentially request these chunks and the supplying peers preferentially utilize their outbound bandwidth for providing these chunks.

#### 4.1.2 *System Architecture*

The system architecture for the MVP2P design is illustrated in Fig. 5. There are four different types of components: tracker, broadcaster, super-peer and peer.

(1) The tracker is a central controller. The broadcaster, super-peers and peers register themselves to the tracker once they join the network. According to the list of registries, a peer can select its neighbouring peers. The tracker stores a Layer Description Table which describes the MVC video layer dependency relationship and the average bit-rate of each layer. When a new peer obtains the above table from the tracker, it knows which layers are required for decoding its observing layer. The tracker also maintains a Peer Status Information Table. The tracker periodically collects the peers' status information (e.g., the outbound bandwidth and their observing layers) to update the above table. It calculates the total peers' outbound bandwidth for each peer subset defined in the maximum flow model and saves this information into a Peer Subset Outbound Bandwidth Table. Once a peer obtains the Layer Description Table and Peer Subset Outbound Bandwidth Table, it can construct the proposed maximum flow model and execute the Ford–Fulkerson algorithm to calculate the maximum flows.

(2) The broadcaster is the streaming originator. It receives the video content from media sources (e.g. live camera or stored video files), creates chunks and distributes the chunks to super-peers. For the chunk encapsulation, the VID and TID are added in the chunk header, so that a peer can request the chunks only for its necessary layers by searching the chunk header.

(3) A super-peer is considered as a streaming server. It has a large outbound bandwidth capacity and stores the entire video contents for all the layers. It receives chunks from the broadcaster and distributes the chunks to the peers.

(4) A peer is an end user. The peers belong to two categories simultaneously: receiving peer and supplying peer. Fig. 6 illustrates the strategies and algorithms which are executed on supplying peers and receiving peers respectively.

When a peer acts as a receiving peer, it executes the chunk scheduling strategy to select the next chunk to be requested. The result of the maximum flow algorithm is used to decide which layer that the next chunk should be requested for. There is an extra step to decide which chunk, which will be explained in Section 4.3. It then executes the peer selection strategy to select the supplying peer to download the selected chunk. The result of the maximum flow algorithm is used to decide which peer subset that the selected chunk should be requested from. There is an extra step to decide which supplying peer, which will be explained in Section 4.4. Thereafter, it requests the selected chunk from the selected supplying peer.

When a peer acts as a supplying peer, it executes the supplying peer bandwidth scheduling algorithm. The result of the maximum flow algorithm is used to evaluate the capabilities of the receiving peers for providing their requested chunks. It selectively sends chunks to the receiving peers with higher capabilities when its outbound bandwidth is limited.

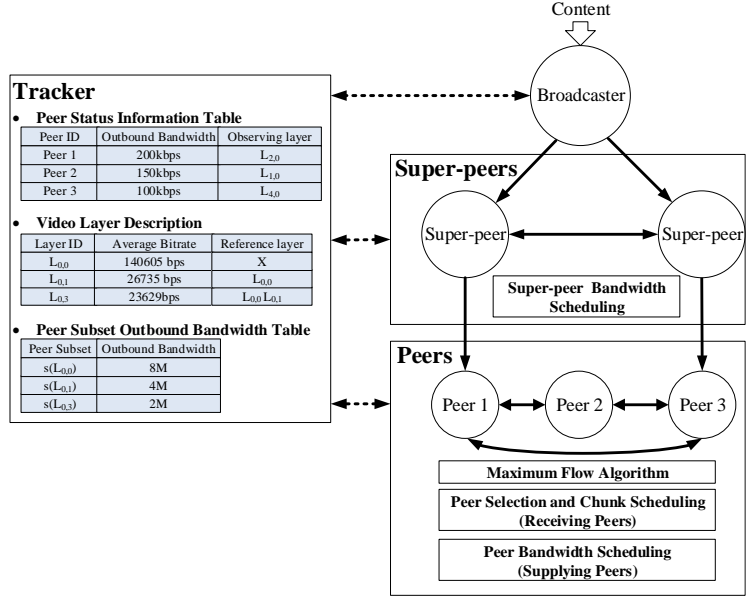

Fig. 5 System architecture

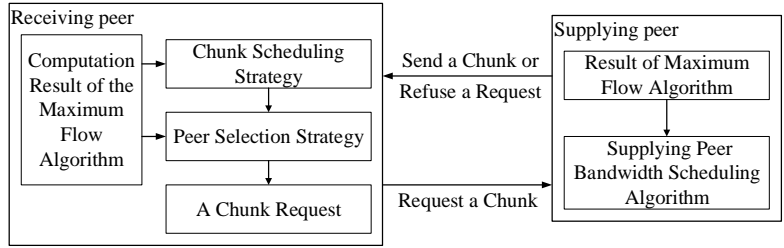

Fig. 6 Strategies and algorithms on supplying peers and receiving peers

## 4.2 Ford–Fulkerson Algorithm for the Maximum Flow Model

In this section, we present the Ford–Fulkerson Algorithm [38]. We construct a complete network model $G(V, E)$ according to the layer dependency, e.g. Fig. 2. Please refer to Fig. 4 for the graph. Let $\{H_1, H_1 …, H_m\}$ denote the $m$ peers watching on layer $L_{i,j}$ and $O(H_k)$ denote the outbound bandwidth of peer $H_k$. For each edge $e(v, s(L_{i,j}))$, the edge capacity is $c(v, s(L_{i,j})) = \sum_{k=1}^{m} O(H_k)$. Suppose $n$ peers require layer $L_{i,j}$, at most $n-1$ peers can obtain layer $L_{i,j}$ from the peers because at least one copy of layer $L_{i,j}$ should be obtained from the servers. Let $R(L_{i,j})$ denote the bitrate of layer $L_{i,j}$. For each edge $e(r(L_{i,j}),t)$, the edge capacity is $c(r(L_{i,j}),t)=(n-1)*R(L_{i,j})$ $(n>0)$.

To compute the maximum flow, denoted by $u(v,t)$, we define the residual network $G_f(V, E_f)$ as the network with capacity $c_f(x,y) = c(x,y)-u(x,y)$ where $x$ and $y$ are two adjacent nodes. The basic idea of Ford–Fulkerson algorithm is as follows. It searches an augmenting path in $G_f(V, E_f)$. An augmenting path is a path from the source $v$ to sink $t$ with available capacity on all edges in the path. As long as there is an augmenting path, it routes flow with a flow value which is equal to the minimum residual edge capacity amongst all edges in the augmenting path. The algorithm complexity is $O(Ef)$ where $E$ is the number of edges in the graph and $f$ is the maximum flow in the graph. There are many other more efficient algorithms (e.g. Edmonds and Karp [38]). However, their implementations are not the focus in the paper. A normal MVC video has 8 views. When it is encoded using the typical Bi-prediction structure as shown in Fig. 1, the number of video layers is 32. For an MVC video with a large number of views (e.g. 100 views), the views are normally grouped into different group size (e.g. 8 views) for encoding. Due to the small number of layers, the workload of the Maximum Flow Model calculation is not high.

In the current design, the maximum flow algorithm is operated on the tracker which sends out the calculated results to all peers. However, the design can be improved using peers to spread the result to reduce the burden of the tracker, which will be a future work of this paper. For the current design, the total transmission delay from the tracker to all peers can be estimated based on $\max(TrackerToPeerRTTs) + MessageSize / \min(TrackerBandwidth/PeerNumber, \min(PeerBandwidths))$. The number of the video layers for a normal MVC video is not large, and consequently the data size for storing and transmitting the result is small. For example, when an MVC video with 5 views is encoded using the typical bi-prediction structure, the number of video layers is 20. Based on Fig. 4, less than 20x20 values need to be stored and transmitted. Suppose each value takes 4 bytes, the tracker's bandwidth is 100Mbps (and it is the bottleneck in transmitting the messages), and the peer number is 1000, for the above video with 20 layers, it will take 16ms plus the network delay to send out the result.

The purpose of the algorithm is to decide the optimal solution for peer bandwidth scheduling. View switching and peer churn will change peer bandwidth distribution. To obtain the optimal solution, the algorithm needs to be re-calculated for every peer bandwidth change. However, when view switching happens frequently, the tracker may calculate the algorithm very frequently, and consequently the workload for the tracker becomes large. To avoid frequent re-calculation, different strategies to re-calculate the algorithm for view switching and peer churns can be applied to achieve an approximate optimal solution, e.g. setting up a threshold (e.g. 15s) for re-calculating the algorithm. If bandwidth changes happen within this threshold, the algorithm is re-calculated periodically based on the threshold, otherwise the re-calculation is triggered by the bandwidth change events. The effects of different re-calculation intervals on the performance are discussed in the Section 5.2.2.

### 4.3 *Chunk Scheduling for Receiving Peers*

#### 4.3.1 *Sliding Window Design*

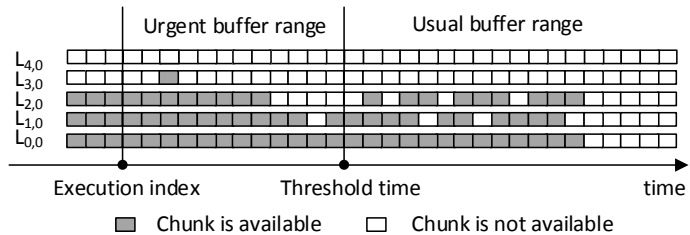

Fig. 7 Chunk buffer management

In MVP2P, each peer maintains a sliding window for each MVC layer as depicted in Fig. 7. The execution index is defined as the next chunk ID to be played. Suppose we use a variable chunk size that each chunk contains the NALUs of each layer for a specified time period (e.g., 1 second), the chunks of different layers are aligned with the time for decoding the video. We specify the threshold time to assess whether a chunk is time-sensitive for the smooth video playback. Consequently, the sliding window of each layer is divided into the "urgent" and "usual" ranges. In the "urgent" range, chunks are time-sensitive. In the "usual" range, chunks don't have strict time constraints.

#### 4.3.2 *Strategy Description*

The chunk scheduling strategy is described as following two parts. A peer can request multiple chunks at a time. The number of requests is based on its inbound bandwidth. The chunk scheduling strategy decides the requesting orders of the chunks for a peer.

#### (1) **Sequential Chunk Scheduling Strategy**

Let us start at dealing with the following three situations that a peer should request chunks to ensure a smooth video playback.

Firstly, when there are missing chunks in the "urgent" range, the missing chunks are time-sensitive. A Sequential Chunk Scheduling Strategy is used. The peer sorts the missing chunks based on the timestamps for decoding the chunks in ascending order. When many chunks have the same timestamp for decoding, the peer sorts the chunks based on layer DONs in ascending order. Thereafter the peer requests them sequentially from lower orders to higher orders. This enables that the chunk nearest to the execution index is downloaded first, and consequently increases the possibilities of a smooth video playback.

Secondly, in the start-up stage, a peer needs a pre-buffering time to buffer some chunks before they start the video playback. This is to enable that the downloading rate can adapt to the bandwidth variations and then the peer can obtain a smooth video playback. The pre-buffer time is different in different video streaming systems. During the pre-buffering time, the peer requests chunks with the above Sequential Chunk Scheduling Strategy.

Thirdly, in the layer switching stage, a peer also needs a buffering time to fetch chunks for new layers. The general human-computer interaction delays should be less than $1s$ as suggested in [39]. ITU recommends that the channel switching delays of an IPTV system should be below $2s$ to guarantee a satisfactory Quality of Experience [40]. This paper adopts $1s$ in the design and performance evaluation. Note that the given buffering time may be longer than the "urgent" range. During the buffering time, the peer requests the chunks for the new layers with the above Sequential Chunk Scheduling Strategy.

(2) **Balanced Chunk Scheduling Strategy**

Except in the above three situations, a peer uses a Balanced Chunk Scheduling Strategy to request chunks for improving chunk availabilities in the network and at the same time ensuring a smooth video playback.

According to the computation result of the maximum flow model proposed in Section 4.2, we can know which layers that each peer subset (supplying node) should provide. The basic idea is that a peer preferentially requests the chunks of the layers that it should provide (**contribution layers**). This improves the chunk availabilities of its contribution layers and then other peers can retrieve the chunks of the contribution layers more easily. Hence, the server bandwidth consumption can be reduced.

As we know, if a peer greedily requests the chunks of the distant future (i.e. the decoding time is far from the current time) for its contribution layers, the chunks of the near future for both contribution layers and non-contribution layers will enter the "urgent" range at a certain time. This increases the possibilities of unsmooth video playbacks. To prevent this, a peer needs to balance the requests for the chunks of contribution layers and the chunks of the near future.

In the following, we present the Balanced Chunk Scheduling Strategy. It strives to enable that a peer consumes enough inbound bandwidth to download near future chunks for smooth video playbacks and consumes the remained inbound bandwidth to download chunks for improving chunk availabilities of contribution layers. The Balanced Chunk Scheduling Strategy combines the above Sequential Chunk Scheduling Strategy and a newly-defined Maximum-Flow-based Chunk Scheduling Strategy.

Let $\sum_{L_{i,j}=L_{0,0}}^{L_{u,v}} R(L_{i,j})$ denote the total bitrate of the required layers for peer $H_k$. It also represents the amount of inbound bandwidth that peer $H_k$ should allocate for downloading its required layers in order to obtain the smooth video playback. Suppose $I(H_k)$ denotes the inbound bandwidth of peer $H_k$, the percentage of the inbound bandwidth that is used to download near future chunks for the smooth video playback is shown below.

$$p(H_k) = \frac{\sum_{L_{i,j}=L_{0,0}}^{L_{x,y}} (R(L_{i,j}))}{I(H_k)} \tag{3}$$

When a peer decides the next chunk to be requested, it first decides whether to request it for the smooth video playback or for improving chunk availabilities of contribution layers based on equation (3). The $p(H_k)$ indicates the probability if the next chunk should be requested for the smooth video playback.

If the next chunk is requested for the smooth video playback, the proposed Sequential Chunk Scheduling Strategy is used. This enables a peer which has just enough inbound bandwidth for downloading its required layers to use the sequential chunk scheduling strategy all the time. Otherwise, a Maximum-Flow-based Chunk Scheduling Strategy is used, which consists of the following two steps.

(i) Layer Selection. This step decides from which layer a peer should request the next chunk. The motivation is to enable a peer to download more chunks for the contribution layer that a peer should allocate more outbound bandwidth. The layer is selected based on the outbound bandwidth contribution when the peer acts as a supplying node. According to the result of the maximum flow model proposed in Section 4.2, for a particular supplying node (e.g. $s(L_{2,0})$), we can get the flow value on each edge from the supplying node to each receiving node (e.g., $r(L_{0,0})$ and $r(L_{2,0})$ in Fig. 4). Let us use $u(s(L_{i,j}), r(L_{x,y}))$ to denote the flow value on edge $e(s(L_{i,j}), r(L_{x,y}))$ and $u(v,s(L_{i,j}))$ to denote the total flow value going through $s(L_{i,j})$. In peer subset $s(L_{i,j})$, the proportion of the allocated peers' outbound bandwidth that is used for providing layer $L_{x,y}$ is shown below:

$$\rho_{r(L_{x,y})} = \frac{u(s(L_{i,j}), r(L_{x,y}))}{u(v, s(L_{i,j}))} \tag{4}$$

A peer selects a layer for downloading based on the probability defined by equation (4). The $\rho_{r(L_{x,y})}$ indicates the probability that a layer $L_{x,y}$ is selected.

(ii) Chunk Selection. The above step only decides which layer for requesting the next chunk. An extra step is required to decide which chunk should be requested. To improve the chunk availabilities of a selected layer, the "rarest-first" chunk scheduling strategy is used where the rarest chunk in the network (i.e. least number of copies in its neighbouring peers) is selected. If there are multiple rarest chunks, the chunk with the smallest chunk ID is selected because it is the most important from the peer's decoding perspective.

### 4.4 *Peer Selection for Receiving Peers*

The peer selection strategy is to select a supplying peer for downloading a chunk. As discussed in Section 4.3, there are three situations where chunks are time-sensitive for smooth video playback. In these three situations, we mark the chunk that is selected to request as "emergent". A Partial Maximum-Flow-based Peer Selection Strategy is employed to decide the supplying peer for downloading "emergent" chunks. For other non-emergent chunks, a Pure Maximum-Flow-based Peer Selection Strategy is proposed to choose the supplying peer.

### (1) **Pure Maximum-Flow-based Peer Selection Strategy**

This strategy is to enable that the peers' outbound bandwidth consumption conforms to the result of the maximum flow model proposed in Section 4.2. It consists of the following two steps.

(i) Peer Subset Selection. This step decides the peer subset that a peer should select for downloading a chunk. According to the result of the maximum flow model proposed in Section 4.2, for a particular receiving node (e.g. $r(L_{4,0})$), we can get the flow value on each edge from each supplying node (e.g. $s(L_{3,0})$ and $s(L_{4,0})$ in Fig. 4) to the receiving node. Let $u(s(L_{x,y}), r(L_{i,j}))$ denote the flow value on the edge $e(s(L_{x,y}), r(L_{i,j}))$ and $u(r(L_{i,j}), t)$ denote the total flow value going through $r(L_{i,j})$. To provide layer $L_{i,j}$, the proportion of the outbound bandwidth of the peer subset $s(L_{x,y})$ with respect to the total is shown below:

$$\rho_{s(L_{x,y})} = \frac{u(s(L_{x,y}), r(L_{i,j}))}{u(r(L_{i,j}), t)} \tag{5}$$

A peer selects a peer subset for downloading a chunk based on the probability defined by equation (5). The $\rho_{s(L_{x,y})}$ indicates the probability that peer subset $s(L_{x,y})$ is selected.

(ii) Peer Selection. The above step only decides which peer subset should be selected. An extra step is needed to decide which peer should be selected. If the selected peer subset $s(L_{x,y})$ contains multiple available peers (that have the chunk and the outbound bandwidth is available), the peer randomly selects an available peer. This avoids the flashing crowd problem, i.e. all peers request a chunk from one peer at the same time. If there are no available peers, the peer requests the chunk from the servers. The servers will refuse most of the requests because the chunk is not emergent, which will be explained in Section 4.5.

### (2) **Partial Maximum-Flow-based Peer Selection Strategy**

When a chunk is marked as "emergent", the Partial Maximum-Flow-based Peer Selection Strategy is used. For the purpose of a smooth video playback, this strategy may not exactly comply with the result of the maximum flow model proposed in Section 4.2.

We initially evaluate the transmission time if the chunk is downloaded from the servers. Let $C$ denote the chunk size, $I(H_k)$ denotes the available inbound bandwidth of the requesting peer $H_k$, and $O(s)$ denotes the available outbound bandwidth of the servers. $RTT$ is the possible maximum round trip time. The formula is shown as below:

$$T_s = RTT + \max\{\frac{C}{I(H_k)}, \frac{C}{O(s)}\} \tag{6}$$

We specify that, once a supplying peer receives the request for an "emergent" chunk, it will immediately decide whether to send the chunk or refuse the request.

In order to increase the probability that an "emergent" chunk can be received on time, the following evaluation is performed. We estimate the duration from a peer sending a request to another peer to a refusal message being received by $T_p = RTT$. Let $T_d$ denote the time to decode the chunk and $T_r$ denote the current time. The following formulation is used to decide if a peer can request the chunk from another peer.

$$J = T_d - T_r - T_s - T_p \tag{7}$$

When $J<0$, it indicates that, if a peer requests a chunk from another peer and the request is refused, it may not have enough time to obtain the chunk from the servers. In this case, the peer will directly request the chunk from a server. Otherwise, the following procedure will be executed. In the procedure, if the peer finds an available peer in any step, it will not execute the rest of the steps.

(i) The Pure Maximum-Flow-based Peer Selection Strategy is used. This step tries to select a supplying peer in terms of the result of the maximum flow model proposed in Section 4.2.

(ii) In this step, the peer subset selection probability is not constrained to equation (5). This step tries to select a supplying peer that is close to the result of the maximum flow model proposed in Section 4.2. All peer subsets that provide the chunk's layer according to the result of the maximum flow model proposed in Section 4.2 are involved. A peer randomly selects a peer which can provide the chunk from these peer subsets with the same probability.

(iii) The peer subset selection is not constrained to the maximum flow model proposed in Section 4.2. This step tries to select a peer for providing the chunk instead of a server. All peer subsets are involved in the selection. The peer randomly selects a supplying peer that can provide the chunk from all the peer subsets with the same probability.

(iv) It resorts to a server. If the all the above steps have failed to find a peer that can provide the chunk, this step chooses the server as the supplier and tries to guarantee the smooth video playback.

### 4.5 Supplying Peer Bandwidth Scheduling

A supplying peer is a server or a normal peer which is currently providing chunks. In BitTorrent-like P2P streaming systems, the receiving peers of a supplying peer don't know what chunks other peers are requesting. The receiving peers compete for a supplying peer's outbound bandwidth, and consequently a supplying peer may receive an excessive number of requests. For example, when a server delivers a new chunk to a peer, a large number of peers may request the chunk from that peer. With the limited outbound bandwidth, the peer may not be able to deliver chunks to all the receiving peers on time. An efficient algorithm for processing these requests is required to deal with the flash crowd events.

In the following, we present the Supplying Peer Bandwidth Scheduling algorithm. For each supplying peer, there is a queue to store requests from the receiving peers. The supplying peer periodically processes the queue. When the flash crowd event happens, it will selectively process the requests. It selectively delivers the chunk to peers with higher capabilities so that the chunk can be disseminated quickly in the network. The algorithm is described in following two conditions.

*(1) The supplying peer is a server.* The algorithm tries to minimize the servers' own bandwidth consumption.

In terms of the result of the maximum flow model proposed in Section 4.2, for each receiving node $r(L_{x,y})$, we know the flow value $u(r(L_{x,y}),t)$ on edge $e(r(L_{x,y}),t)$. We also know the edge capacity $c(r(L_{x,y}),t)$ of edge $e(r(L_{x,y}),t)$, which represents the total demand of the receiving node of layer $L_{x,y}$. Let $R(L_{x,y})$ denote the average bitrate of layer $L_{x,y}$. To fully provide layer $L_{x,y}$, the outbound bandwidth that the servers should allocate, denoted by $Q_s(L_{x,y})$, has been shown in equation (2).

Based on equation (2), we can calculate the number of copies that the server should provide for each chunk of layer $L_{x,y}$, denoted by N. The formulation is shown below.

$$\text{N} = \text{ceil}(\frac{Q_s(L_{x,y})}{R(L_{x,y})}) \tag{8}$$

As discussed in Section 4.4, a peer will directly request a chunk from a server when $J < 0$ in equation (7). Such requests have the highest priorities to be processed by the server. When the server receives such a request, it will send the chunk to the receiving peer immediately. If the server bandwidth is insufficient to process such a request, that means the total network bandwidth is insufficient for serving global peers in the network. The system should cut down the number of peers or increase the server capacities.

For other types of chunk requests, i.e. $J \geq 0$ in equation (7), the server tries to maximize the capabilities of the receiving peers for providing their requested chunks so that the server's bandwidth consumption is minimized. The server will check the number of copies for the chunk that it has sent out, denoted by $M$. If $M < N$ and the receiving peer will supply the chunk to other peers according to the result of the maximum flow model, the server will execute the following steps.

According to equation (4), each receiving peer only allocates a particular ratio of outbound bandwidth for providing a contribution layer while acting as a supplying node. Let $O(H_k)$ denote the outbound bandwidth of a receiver peer $H_k$. $\rho_{r(L_{i,j})}$ in equation (4) denotes the proportion of the outbound bandwidth that $H_k$ allocates for providing layer $L_{i,j}$. We use the outbound bandwidth that $H_k$ allocates for providing layer $L_{i,j}$ to represent $H_k$'s capability for providing its requested chunk "$c$" of layer $L_{i,j}$ It is formulated as below.

$$\text{A}(L_{i,j}, c, H_k) = \text{O}(H_k) \times \rho_r(L_{i,j}) \tag{9}$$

The server sorts the requests based on the receiving peers' capabilities in descending order. If the receiving peers' capabilities for some requests are the same, the requests are sorted based on the rareness of the requested chunks (i.e. the number of copies for the chunk in the network) in ascending orders. The server will always process the request arranged in the head of the queue if and only if $M < N$ and the receiving peer will supply the chunk that it is requesting. Otherwise, the server will refuse the request.

(2) The supplying peer is a normal peer. The algorithm tries to maximize the peer's outbound bandwidth consumption.

When a supplying peer receives a request containing a chunk that is in the "urgent" range, the peer will immediately decide whether to send the chunk or refuse the request based on its available outbound bandwidth, so that the receiving peer has more time to request the chunk from another peer.

When a peer receives a request containing a chunk that is in the "usual" range, the peer tries to maximize the receiving peers' capabilities for providing their requested chunks with its limited outbound bandwidth. The peer will execute following steps:

Firstly, the requests are sorted based on the receiving peers' capabilities in descending order (i.e. A$\left(L_{i,j}, c, H_k\right)$ of equation (9)) and the rareness of the requested chunks in ascending order; Secondly, the peer always processes the request which is arranged in the head of the queue until its outbound bandwidth is completely consumed; Thereafter, it will refuse unprocessed requests, and the rejected peers will send requests to other peers. This refusal process is to implement the Snow-ball process, i.e. the peers with higher contributed outbound bandwidth to the layer are served earlier so that more peers can be served in the future stages.

## 5. PERFORMANCE EVALUATION

In this section, we evaluate the performances of the proposed MVP2P system by comparing it with the existing widely used method which is named as SRT in the paper (i.e. Sequential, Rarest-first, and Tit-for-tat). SRT adopts a sequential chunk scheduling when there are missing chunks in the "urgent" range and a "rarest-first" chunk scheduling strategy when all the chunks in the "urgent" range have been downloaded. SRT employs the tit-for-tat peer selection strategy.

The simulation is built on PeerSim [41] integrated with a BitTorrent module [42]. A random topology generator is used. The peers are linked randomly. The link delay between two peers is randomly generated from 10ms to 300ms. The peer numbers, the inbound and outbound bandwidth of each peer are manually specified in the configuration file as per different experiment requirements.

In all the experiments, the test video sequence "ballroom" [43] is used with the resolution *640x480* and *5* views. The frame rate is *25 fps*. The "ballroom" sequence is encoded by the JMVC reference software [44] with the typical Bi-prediction structure (as shown in Fig. 1) [3]. The GOP size is *8*. An Instantaneous Decoder Refresh (IDR) picture is inserted every *3* GOPs (around *1* second). The length (L) of the "ballroom" sequence is *10* seconds. The data size (S) of NALUs for each layer are separately calculated. The average bitrate (B) of a layer is $S/L$. The average bit-rate of each layer is shown in TABLE I.

Each experiment is set for *600 s*. Due to the short length of the "ballroom" sequence, the broadcaster repeatedly reads the video stream. For each layer, the broadcaster encapsulates the NAULs of *3* GOPs starting from an IDR access unit into a chunk, i.e. one chunk contains the NALUs of a layer for *1 s*. Consequently, an IDR picture is contained in each chunk of the base layer. Each chunk of the base layer provides a random access point from which a new peer can start to download chunks and decode the video stream correctly.

For each peer, the start-up delay (pre-buffering time) is set to *6 s*. To enable all the peers to have a similar video playback position, we assume that there is a virtual media player in the system and it starts playing video after the broadcaster creates chunks for *30 s*. The tracker records the current execution index of the virtual media player. Each new peer obtains the above execution index from the tracker, and afterwards, the peer selects its own random access point which should be *6 s* later than the above execution index. When the peer has received chunks for *6 s*, it starts the video playback. The length of the chunk buffer is assumed to be infinite. The "urgent" chunk buffer range is set to *4 s*. Each peer processes its request queue every *200 ms*.

The peers join the network at random times from the *30s* (i.e., after the virtual media player starts playing video) to *100 s*. Their observing layers are randomly chosen. To easily evaluate the MVP2P method, we analyse the experiment results after all the peers have joined.

TABLE I

The average bit-rate (BPS) of each layer of the encoded video "ballroom" (Note: V represents View and T represents temporal layer)

| | T0 | T1 | T2 | T3 |
|---|---|---|---|---|
| V0 | 268323 | 62919 | 82825 | 98368 |
| V1 | 147404 | 43208 | 65198 | 90268 |
| V2 | 204022 | 62569 | 81888 | 99579 |
| V3 | 139005 | 41132 | 64350 | 90092 |
| V4 | 223238 | 69182 | 91279 | 110801 |

### 5.1 Evaluation of System Scalability

The server bandwidth consumption of a P2P streaming system indicates the scalability performance of the system design. If peers consume less server bandwidth, the remaining server bandwidth can be used to serve more peers. This section compares MVP2P with SRT under different peer bandwidth settings and peer numbers. In the experiments, the start-up delay is set to *6s*. In both methods, if a peer doesn't receive a chunk at a certain time (i.e. $J < 0$ in equation (7)), it will request the chunk from the servers. The bandwidth capacity of the servers is assumed to be unlimited in the experiments for evaluation purposes. The server bandwidth consumption reflects the efficiency of the methods.

#### 5.1.1 *Test Results for Different Peer Bandwidth Settings*

##### 5.1.1.1 *Homogeneous Peer Bandwidth Settings*

According to TABLE I, the average bitrate of the "ballroom" sequence is *2,135,650 bps*. In order to ensure that each peer can watch on its randomly selected layer, the inbound bandwidth of each peer is set to *2 Mbps*. Different bandwidth conditions are presented through setting the ratio of each peer's outbound/inbound bandwidth. The peer number is set to *100*.

Let "*S*" denote the amount of server bandwidth consumption and let "*U*" denote the total amount of bandwidth consumption for serving the global peers in the network. The percentage of server bandwidth consumption is formulated as below.

$$P_s = \frac{S}{U} \tag{10}$$

Fig. 8 exhibits the percentage of server bandwidth consumption for MVP2P and SRT. For comparison, the theoretically optimal value based on the maximum flow model proposed in Section 4.2 is also present.

When the ratio of outbound/inbound bandwidth is *0*, no peer contributes to the network. We calculate the total bandwidth consumption for serving the global peers. It is *73,254,988 bps* for MVP2P, SRT and the theoretical optimal value.

When the ratio of each peer's outbound/inbound bandwidth is equal to or greater than *0.4*, the peers' outbound bandwidth is sufficient to serve the global peers. The percentage of server bandwidth consumption for the MVP2P method is *3%*, which is the same as the theoretical optimal value. By contrast, the percentage of server bandwidth consumption for the SRT method keeps steady at around *80%*. MVP2P takes full advantage of the peers' outbound bandwidth for providing layers. In SRT, peers form a streaming structure where peers cannot efficiently share their layers. The peers' outbound bandwidth is underutilized. This is the main reason that the server bandwidth consumption is still huge although the peers' outbound bandwidth becomes more abundant.

When the ratio of each peer's outbound/inbound bandwidth is *0.2*, the peers' outbound bandwidth is insufficient to serve the global peers. The percentage of server bandwidth consumption for the theoretical optimal value, MVP2P and SRT is *45%*, *51%* and *81%* respectively. The theoretical optimal value is calculated without considering the chunk dissemination delay. Since the peers' outbound bandwidth becomes smaller, the chunk dissemination delay becomes larger. In this situation, the MVP2P consumes extra server bandwidth to ensure that chunks are delivered to peers before the required decoding deadline.

##### 5.1.1.2 *Heterogeneous Peer Bandwidth Settings*

In [47], the proportions of the global broadband adoptions with the average connection speeds above *4 Mbps*, *10 Mbps*, 15 Mbps and *25 Mbps* are *82%*, *46%*, *28%* and *12%* respectively. The average bitrate of the test video sequence in this paper is *2,135,650 bps*. For broadband uses with the average connection speed below *4 Mbps*, they must have the average connection speed above *2 Mbps* in order to watch the test video smoothly. Therefore, in the experiments, the distribution of the peer bandwidth is set as follows; the percent of peer inbound bandwidth with *2 Mbps*, *4 Mbps*, *10 Mbps*, *15 Mbps* and *25 Mbps* is *18%*, *37%*, *17%*, *16%* and *12%* respectively. The peer number is set to *100*.

Free riders can be eliminated using some incentive policies, e.g. TBeT [48]. MVP2P and the incentive policies are complementary to each other. Free riders do not affect the operation of MVP2P, which are simply considered as contributing zero bandwidth to the network. In the experiments, *5%* peers are set as free riders. The other experiment parameters are the same as the above experiments (the homogeneous peer bandwidth settings).

The results are shown in Fig. 9. Much similar to the results in Fig. 8, MVP2P still performs better than SRT. MVP2P calculates the Maximum Flow Algorithm based on the global peers' outbound bandwidth. Even though some peers do not contribute their outbound bandwidth, the heterogeneous peer bandwidth settings have no impact on the effectiveness of MVP2P.

When the ratio of each peer's outbound/inbound bandwidth is *0.2*, the percentage of server bandwidth consumption for MVP2P is *3%* (as shown in Fig. 9) whereas it is *51%* in Fig. 8. This is because the total peer outbound bandwidth becomes sufficient because of the larger broadband adoptions.

#### 5.1.2 *Test Results for Different Peer Numbers*

We conduct two groups of experiments: small peer number group; large peer number group. The peer number varies from *10* to *90* in the small peer number group tests. The peer number varies from *100* to *500* in the large peer number group tests. In all the experiments, the inbound bandwidth of each peer is *2 Mbps* and the ratio of each peer's outbound/inbound bandwidth is set to *0.4*.

The tests are shown in Fig. 10 and Fig. 11. In both figures, MVP2P performs much better than SRT.

In Fig. 10, when the peer number is *10*, the percentage of server bandwidth consumption for MVP2P is *24%* whereas it is *83%* for SRT. In this case, the total bandwidth consumption for serving the global peers is *5,592,437 bps*. The MVP2P method achieves the theoretical optimal value where only one copy of each required layer is supplied by the servers. In this situation, the sever bandwidth consumption is *1,371,003 bps*, which is less than the average bitrate of the "ballroom" sequence. This is because that the required layers for all the randomly generated peers don't cover all the layers.

When the peer numbers increase, the percentage of server bandwidth for the MVP2P method declines whereas the percentage of server bandwidth for the SRT method keeps steady with a high server bandwidth consumption. According to equation (10), "$U$" increases when the peer number increases. Since the MVP2P method still achieves the theoretical optimal value, "$S$" keeps the same. Therefore, "$P_s$" for MVP2P drops. In SRT, since the server bandwidth consumption is not improved, "$S$" also increases. "$P_s$" of SRT method is still *80%*.

In Fig. 11, when the peer number becomes larger, the curve of MVP2P continually drops while the curve of SRT keeps stable at around *80%*. When the peer number is *500*, the server bandwidth for the MVP2P method is *1,990,872 bps*. By contrast, the server bandwidth for the SRT method is up to *289,789,270 bps* which is approximate *135* times of the average bitrate of the "ballroom" sequence. When the peer number becomes larger, the flash crowd events are more likely to happen. In MVP2P, the supplying peer bandwidth scheduling algorithm can efficiently deal with flash crowd events. SRT doesn't consider this situation.

### 5.2 Evaluation of View Switching

View switching happens frequently in free viewpoint television scenarios. In this section, we assume that the servers' bandwidth capacity is unlimited for evaluation purposes. When $J < 0$ in equation (7), a peer will request the chunk from a server and we presume that the server can always deliver the chunk to the peer on time. Consequently, a peer can always successfully switch to its new observing layer in the view switching stage. With this setting, the view switching delay is reflected by the server bandwidth consumption. We only need to evaluate the server bandwidth consumption for MVP2P and SRT.

#### 5.2.1 Different Layer Switching Rates

The additional experiment parameters are as follows. The peer number is *100*. The inbound bandwidth of each peer is *2 Mbps* and the ratio of each peer's outbound/inbound bandwidth is *0.4*. The layer switching delay is set to *1 s* as per the switching delay requirement discussed before. Random layer switching pattern is used. The layer switching time for a peer is randomly generated and inserted into the simulation. Since the simulation is event driven, it always executes layer switching events in time orders. In other words, even if some peers have the same layer switching time, they don't switch simultaneously. The layer switching rate is specified for each experiment. For example, if the layer switching rate is set to *1 time/min*, it represents that each peer will execute a layer switching event in a minute on average. Due to layer switching events, the outbound bandwidth of supply nodes (in the maximum flow model) changes. In MVP2P, the maximum flow model is re-calculated corresponding to the layer switching rate (e.g. *1 time/min*).

Fig. 12 shows the percentage of server bandwidth consumption for MVP2P and SRT. With respect to Fig. 11 the actual amount of server bandwidth consumption for both methods is shown in Fig. 13. As we can see, MVP2P outperforms SRT significantly. When the layer switching rate is up to *20 times/min*, the percentage of the server bandwidth consumption for the MVP2P method is *6%* as shown in Fig. 12. The actual server bandwidth consumption is *7,359,141 bps* in Fig. 13, which is only *3.5* times more than the average bitrate of the "ballroom" sequence. By contrast, the percentage of the server bandwidth consumption for SRT is *69%* in Fig. 12, and the actual server bandwidth consumption is *78,560,495 bps* in Fig. 13, which is *36.7* times more than the average bitrate of the "ballroom" sequence.

When the layer switching rate increases, the percentage of server bandwidth consumption for the MVP2P method slightly increases. When a peer switches to a new observing layer, it may not have the chunks of the new layers for the next video playback. According to equation (7), a peer can request a chunk from another peer if and only if $T_d - T_r \geq T_s + T_p$. In the experimental settings, the link delay is from *10 ms* to *300 ms*. The possible maximum RTT is *600 ms*. Hence, $T_s + T_p$ is greater than 1.2s. As the maximum layer switching delay is set as *1s*, i.e. $T_d - T_r = 1s$, the peer will directly request the chunks from servers in this situation. This consumes extra server bandwidth. The theoretical optimal value is calculated without considering the link delays. In a real P2P network environment, it is difficult to achieve the optimal value especially when the link delay is large.

In Fig. 12, when the layer switching rate increases, the percentage of server bandwidth consumption for SRT decreases. In fact, the actual server bandwidth consumption increases as shown in Fig. 13. In these cases, the amount of the total amount of bandwidth consumption for serving the global peers grows quicker than the amount of the server bandwidth consumption. When the layer switching rate is high (e.g. *15 times/min*), a peer frequently switches the observing layer and consequently it needs to download the chunks of almost all the layers.

#### 5.2.2 Different Maximum Flow Algorithm Calculation Intervals

In above experiments, the Maximum Flow Algorithm is re-calculated corresponding to the layer switching rate. For example, when the layer switching rate is *1 time/min*, the Algorithm is recalculated every *60 seconds*.

In this set of experiments, the layer switching rate is fixed to *5 times/min*. However, different calculation periods (or intervals) are given for the Maximum Flow Algorithm. The other parameters are the same.

The results are shown in Fig. 14 and Fig. 15. As SRT doesn't use the Maximum Flow Algorithm, its performances remain the same. However, the changes in MVP2P performances are noticeable.

When the Maximum Flow Algorithm interval is *5 s*, the percentage of server bandwidth consumption for MVP2P is *7%* and the actual server bandwidth consumption is *7,354,712 bps*. As the interval is short, the times that peers re-schedule their bandwidth may become more. The re-scheduling actually changes the relationship between supplying peers and receiving peers. Some supplying peers may need some buffering time to download chunks and their receiving peers may not retrieve chunks in time. This may consume more server bandwidth.

When the Maximum Flow Algorithm interval is *15 s*, the performance of MVP2P is closest to the theoretical optimal value. The percentage of server bandwidth consumption for MVP2P is 4% and the actual server bandwidth consumption is *4,433,314 bps*. When a new result of the Maximum Flow Algorithm is calculated, peers needs to form a new relationship between supplying peers and receiving peers. Based on the experiment results, it takes around *10 s* for peers to form the relationship so that their bandwidth-scheduling plan comply with the result of the Maximum Flow Algorithm. In the following *5 s*, the server bandwidth consumption is close to the theoretical optimal value.

When the Maximum Flow Algorithm interval is *20 s* or *25 s*, the performances of MVP2P become worse. The layer switching rate is *5 times/min*. In a time period of *12 s*, almost all peers switch to a new layer. Therefore, the old result of the Maximum Flow Algorithm is ineffective *12 s* after it is calculated. The ineffective peer bandwidth scheduling consumes more server bandwidth. A future work is to find the best re-calculation interval based on the layer switch patterns.

### 5.3 *Peer Departures*

This section evaluates the performances of MVP2P and SRT in the networks with peer departures. When a peer leaves the network, it stops contributing chunks to its receiving peers. This may affect the streaming qualities of the receiving peers. In this section, we assume that the servers' bandwidth capacity is unlimited for evaluation purposes, i.e. when $J < 0$ in equation (7), the receiving peers can always successfully retrieve the chunks from the servers. With this setting, the system performance, e.g. delays in waiting for downloading the required layers caused by peer departures, is reflected by the server bandwidth consumption, and therefore we only need to evaluate the server bandwidth consumption for comparing the methods.

The additional experiment parameters are as follows. The peer number is *100*. The inbound bandwidth of each peer is *2 Mbps* and the ratio of each peer's outbound/inbound bandwidth is *0.4*. We add peer departure events into the simulation. A peer departure rate is specified for each experiment. It represents the percentage of the peers that will leave the network in the experiment. The departure time for a peer is randomly generated. Due to peer departures, the outbound bandwidth of the supplying nodes (in the maximum flow model) changes. MVP2P re-calculates the maximum flow model when a peer leaves the network.

Fig. 16 shows that the percentage of the server bandwidth consumption for MVP2P is much lower than that of SRT. MVP2P nearly achieves the theoretical optimal value. According to Section 4.3, peers download the chunks for their contribution layers using the rarest-first chunk scheduling strategy. This enables the chunks of a layer to be distributed evenly amongst the peers that should supply them. When a peer leaves the network, the chunks of its contribution layers are still available in other peers. Its receiving peers can retrieve the chunks from new supplying peers. Furthermore, the maximum flow model is re-calculated when a peer leaves the network. Once the peers obtain the new calculation result, they can rapidly adapt the new peer selection strategy, chunk scheduling strategy and supplying peer bandwidth scheduling algorithm. Therefore, MVP2P is robust to peer departures.

### 6. Conclusion

This paper has proposed a maximum flow based model to minimize the server bandwidth costs in live MVC video streaming scenarios using P2P networks. This model considered the MVC video layer dependency relationship and the layer supplying relationship among peers. A layer dependency aware MVC video streaming method over a BitTorrent-like P2P network is proposed, called MVP2P. MVP2P was compared with the existing methods, e.g. rarest first chunk downloading and tit-for-tat based peer selection. The experimental results show that MVP2P offers a better system scalability than the existing methods, especially when the peer number is large. For example, when the peer number was *500*, the existing method consumes the server bandwidth which is approximate *135* times of that in MVP2P. MVP2P also exhibits a better robustness against view switching and peer departures compared with the existing methods.

A possible challenge in MVP2P is the communication costs for maintaining up-to-date peers' status when the peer number grows. In the future work, we will investigate methods to improve the performance of distributing the Maximum Flow results and the effects of more complex layer switching patterns on the performance.

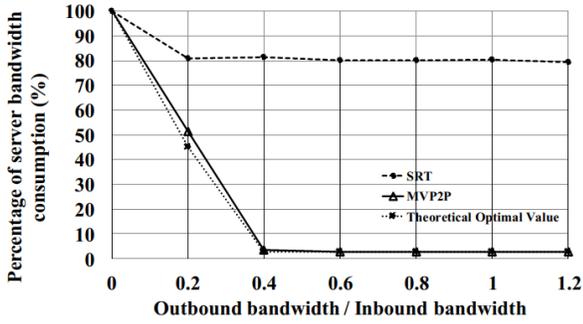

Fig. 8 Comparison of the percentages of server bandwidth consumption with different homogeneous bandwidth conditions

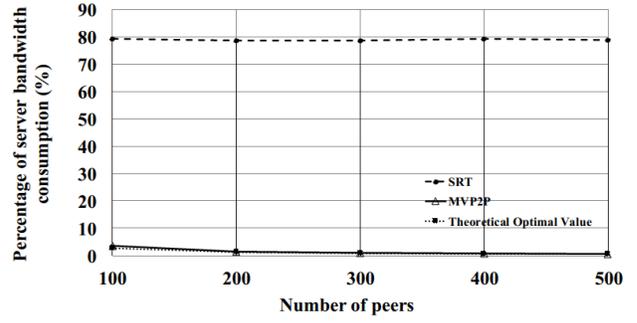

Fig. 11 Comparison of the percentages of server bandwidth consumption with different peer numbers (large peer number)

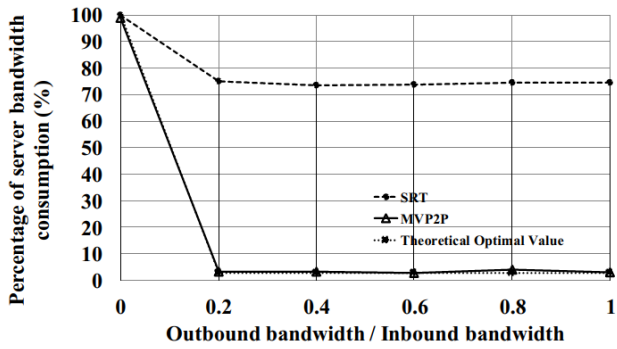

Fig. 9 Comparison of the percentages of server bandwidth consumption with different heterogeneous bandwidth conditions

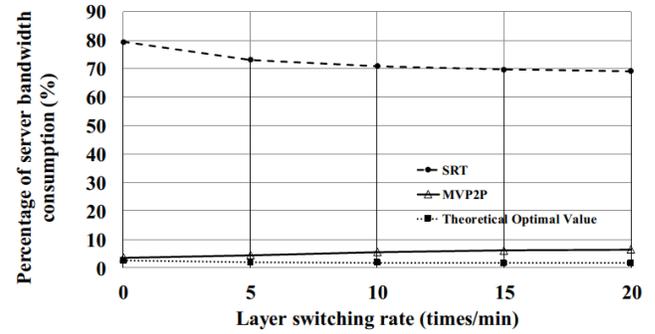

Fig. 12 Comparison of the percentages of server bandwidth consumption with different layer switching rates

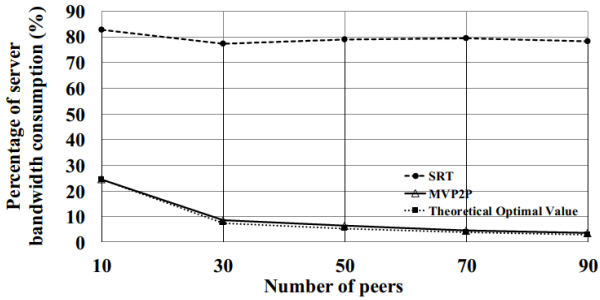

Fig. 10 Comparison of the percentages of server bandwidth consumption with different peer numbers (small peer number)

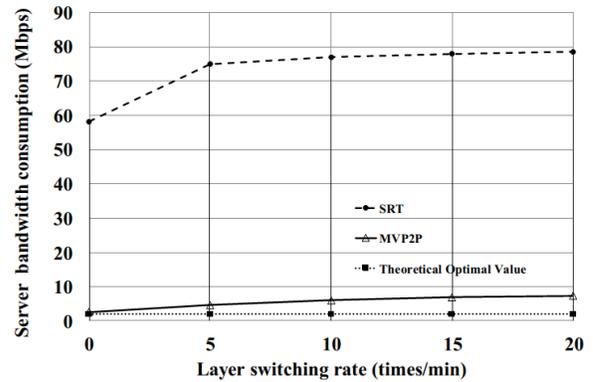

Fig. 13 Comparison of the server bandwidth consumption with different layer switching rates

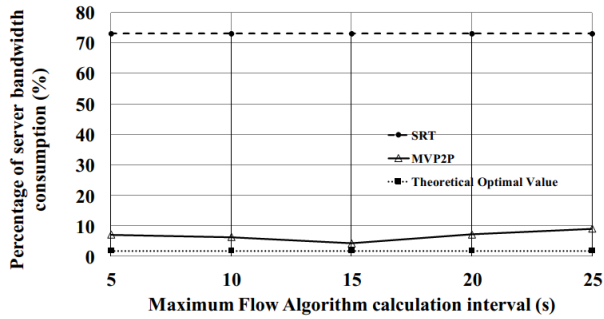

Fig. 14 Comparison of the percentages of server bandwidth consumption with different Maximum Flow Algorithm calculation intervals

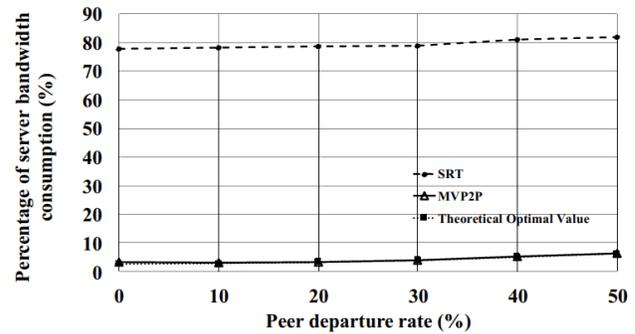

Fig. 16 Comparison of percentage of server bandwidth consumption with different peer departure rates

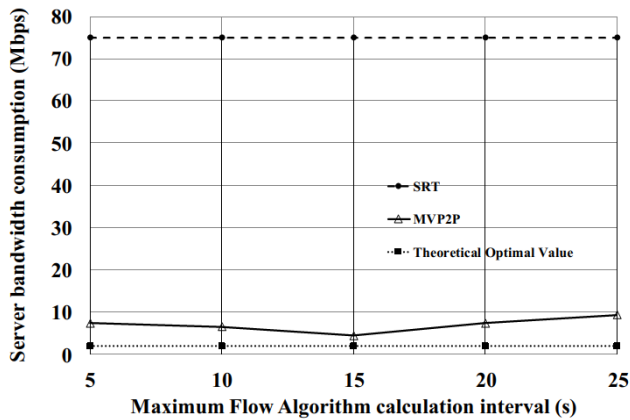

Fig. 15 Comparison of the server bandwidth consumption with different Maximum Flow Algorithm calculation intervals

## ACKNOWLEDGEMENT

This publication has emanated from research supported by a research grant from Science Foundation Ireland (SFI) under Grant Number 13/SIRG/2178, and the AIT President's Seed Fund.